\documentclass[a4paper, 11pt, twoside]{article}
\usepackage[utf8]{inputenc}
\usepackage{comment}
\usepackage{amsmath}
\usepackage{amsfonts}
\usepackage{amssymb}
\usepackage{siunitx}
\usepackage{listings}
\usepackage{graphicx}
\usepackage{amsthm}
\usepackage{enumerate}
\usepackage{listings}
\usepackage{verbatim}
\usepackage{hyperref}
\usepackage{array,epsfig}
\usepackage{amsxtra}
\usepackage{mathrsfs}
\usepackage{hhline}
\usepackage{color}
\usepackage{float}
\usepackage{caption}
\usepackage{subcaption}
\usepackage{bm}
\usepackage{mathtools}
\usepackage{booktabs}
\usepackage{titlesec}
\usepackage[margin=72pt, top=60pt, bottom=105pt]{geometry}
\usepackage[dvipsnames]{xcolor}
\usepackage[numbers]{natbib}

\def\modelname{hierarchical Bayesian model of subpopulation dynamics}
\def\modelnameplural{hierarchical Bayesian models of subpopulation dynamics}

\def\Npatients{150}
\def\Ntestpatients{150}
\def\Nmeasurements{14}
\def\Ndays{392}
\def\noiselevel{1}

\newif\ifRandomEffects
\RandomEffectsfalse

\newif\ifshowfigures
\showfigurestrue
\newif\ifredpart
\redparttrue

\newcommand{\R}{\mathbb{R}}

\newcommand{\vect}[1]{\ensuremath{\boldsymbol{\mathbf{#1}}}} 
\newcommand{\matr}[1]{\ensuremath{\boldsymbol{\mathbf{#1}}}}

\usepackage{authblk}        

\title{Prediction of cancer dynamics under treatment \\ using Bayesian neural networks: A simulated study}

\author[1]{Even Moa Myklebust}
\author[1]{Arnoldo Frigessi}
\author[2,3]{Fredrik Schjesvold}
\author[4]{\\Jasmine Foo}
\author[5]{Kevin Leder}
\author[1, 6]{Alvaro Köhn-Luque}

\affil[1]{Oslo Centre for Biostatistics and Epidemiology, Faculty of Medicine, University of Oslo, 0372 Oslo, Norway}
\affil[2]{KG Jebsen Center for B-Cell Malignancies, Institute for Clinical Medicine, University of Oslo, 0450 Oslo, Norway}
\affil[3]{Oslo Myeloma Center, Department of Hematology, Oslo University Hospital, 0450 Oslo, Norway}
\affil[4]{Institute for Mathematics and its Applications, School of Mathematics, University of Minnesota, Minneapolis, MN 55455, USA}
\affil[5]{College of Science and Engineering, University of Minnesota, Minneapolis, MN 55455, USA}
\affil[6]{Oslo Centre for Biostatistics and Epidemiology, Oslo University Hospital, 0372 Oslo, Norway}

\date{}
\begin{document}

\maketitle

\begin{abstract}
Predicting cancer dynamics under treatment is challenging due to high inter-patient heterogeneity, unclear importance of available biomarkers, and sparse and noisy longitudinal measurements. 
Mathematical models of longitudinal measurements can summarize cancer dynamics by a few interpretable model parameters.
By using machine learning methods to predict the model parameters using baseline covariates, the relationships between the covariates and cancer dynamics under treatment can be discovered. 
However, this two-step approach does not account for the uncertainty of model parameter estimates in the first step. 
Instead, hierarchical Bayesian modeling can be used to model the relationships from baseline covariates to longitudinal measurements through mechanistic parameters while incorporating the uncertainty of each part of the model.

The mapping from baseline covariates to model parameters can be modeled in several ways. 
A linear mapping simplifies inference but fails to capture nonlinear covariate effects and scale poorly for interaction modeling when the number of covariates is large. 
In contrast, Bayesian neural networks (BNNs) can potentially discover interactions between covariates automatically, but at a substantial cost in computational complexity.

In this work, we develop a \modelname{} that uses baseline covariate information to predict cancer dynamics under treatment, inspired by cancer dynamics in multiple myeloma (MM). 
In MM, the level of a specific monoclonal protein (M protein) in a patient's blood is a well-known proxy of tumor burden. 
As a working example, we apply the model to a simulated dataset, comparing the model's ability to predict M protein trajectories to a hierarchical Bayesian model with linear covariate effects.
Our results show that the BNN covariate effect model predicts cancer dynamics under treatment more accurately than a model with linear covariate effects when covariate interactions are present.
This framework is also applicable to other types of cancer or other time series prediction problems that can be described with a parametric model. 
\end{abstract}

\newpage
\section{Introduction}
Predicting cancer dynamics under treatment \cite{bottino2019development, otani2022mathematical, stein2008tumor, wilkerson2017estimation, maitland2020enhanced, browning2024predicting, myklebust2024relapse} is an essential step on the way to personalized cancer therapy. 
It is made challenging by sparse and often noisy longitudinal measurements, often with missing data and for a limited number of patients, with different medical histories and potentially  undergoing different treatments. 
Furthermore, for many diseases, predictive biomarkers are not known but may be discovered from high-dimensional sets of covariates measured with next-generation sequencing techniques. 
This calls for mathematical models that can handle observation noise, missing data, and patient heterogeneity, and discover predictive covariate effects automatically. 

Multiple myeloma (MM) is a heterogeneous cancer with few known predictive biomarkers. 
It is a hematologic malignancy affecting approximately 150,000 people annually worldwide \cite{zhou2021measuring}. 
Prognostic biomarkers are practically non-existing in MM, with the notable exception of the chromosomal translocation t(11;14), which is a predictive biomarker for BCL-2 therapy \cite{bal2022multiple}. 
Consequently, there is a strong need to identify markers of treatment response which can inform treatment decisions \cite{dimopoulos2021multiple}.
MM patients typically have an elevated presence of monoclonal antibodies known as \textit{monoclonal protein} (M protein) in their blood. 
Time series of M protein measurements provide an opportunity to study the underlying population dynamics using statistical inference methods, and has inspired several mathematical modeling approaches in the past \cite{swan1977optimal, tang2016myeloma}. 

Different types of mathematical models from different mathematical fields are often treated separately without any connections. 
However, combinations of methods from different fields offer a huge potential to fulfill various modeling goals at once. 
For example, machine learning is a great way to learn patterns from large datasets, using very flexible methods to discover important covariates automatically. 
On the other hand, machine learning models often lack direct mechanistic interpretation, and quantifying the uncertainty of their predictions typically requires additional post-processing steps. 
As an alternative, hierarchical Bayesian modeling offers statistical uncertainty quantification inherently but requires case-specific model development and careful choices of priors and sampling algorithms. 

We developed a novel hybrid model combining machine learning and statistical uncertainty quantification to benefit from the strengths of both approaches. 
This model combines the uncertainty quantification of Bayesian inference, the interpretability of mechanistic models, and the automatic covariate effect discovery of machine learning in the same framework. 
Our \modelname{} models the relationship between baseline covariates and population dynamics of sensitive and resistant subpopulations in MM through mechanistic model parameters. 
The model uses a nonlinear mixed effect model (NLME) \cite{lindstrom1990nonlinear} framework to share information between patients, and sparsity-inducing priors to combat overfitting.
The model uses a BNN \cite{mackay1992practical} to model the mapping from baseline covariates to mechanistic model parameters, and enables automatic discovery of covariate interactions even for large numbers of covariates. 
In this way, we are able to provide individual predictions of cancer dynamics under new treatment by learning from other patients.

Provided with strongly informative covariates, our model could predict treatment response prior to start of therapy. 
As an example, we applied this model to simulated MM patient data, comparing the method to an alternative method with a linear mapping from covariates to mechanistic model parameters. 
If successful, the model could be used as an aid in choosing the next treatment from a range of candidate treatments. 
More generally, the model framework could be applied to other cancer types or time series prediction problems that can be described with a parametric model.

\section{Materials and methods}
In this section, we describe a mathematical model that predicts cancer dynamics under treatment using covariates measured before the start of treatment, and describe a simulation study designed to investigate whether the model can benefit from a neural network mapping from covariates to model parameters to capture interactions between covariate effects. 

\subsection{Mathematical model of cancer dynamics under treatment}
A response of a multiple myeloma patient to treatment is, in most cases, seen as a decrease in the serum M protein level \cite{kumar2016international}. 
If the patient stays on the same treatment, disease progression typically occurs in the form of increasing M protein level. 
These M protein dynamics can be explained using two broad phenotypes of cancer cells: 
cells that are sensitive to treatment and therefore decrease in amount, and cells that are resistant to treatment and therefore increase in amount. 
The observed M protein value will be the total contributions from these two phenotypic subpopulations.
Mathematically, the total M protein level at time $t$ can be expressed as:
\begin{equation}
\begin{aligned}
    M(t) &= M^0 \pi^r \exp\Big(\rho^r t\Big) + M^0 (1-\pi^r) \exp \Big(\rho^s t\Big), \\
\end{aligned}
\label{eq:model1}
\end{equation}
where $M^0$ is the M protein level at the start of treatment, $\pi^r$ as the proportion of resistant cells at the start of treatment, $\rho^r$ is the growth rate of resistant cells, and $\rho^s$ is the decay rate of sensitive cells. 
Here, it is assumed that each population secretes M protein at the same rate, and that this rate does not change over time. 

Models of this kind have previously been used to model cancer dynamics under treatment, using M protein as a marker of tumor size in multiple myeloma\cite{bottino2019development, otani2022mathematical}; using PSA as a marker of tumor volume in prostate cancer \cite{stein2008tumor, wilkerson2017estimation}; and using tumor volume measurements from CT imaging in colorectal cancer \cite{maitland2020enhanced}. 
Bayesian approaches with different mechanistic models have also been used to predict tumor volume measurements from CT imaging in head-and-neck cancer \cite{browning2024predicting}. 
By parametrizing the decay rate, growth rate, and respective proportion of each subpopulation, future M protein values can be predicted, and causes of disease progression can be explored by quantifying associations between baseline covariates and the model parameters.
The model parameters in equation \eqref{eq:model1} deterministically parametrize an M protein trajectory. 
The influences of each parameter $\rho^r$, $\rho^s$, $\pi^r$, and $M^0$ on the M protein trajectory are shown in Figure \ref{fig:example}. 

\ifshowfigures
\begin{figure}[htbp]
   \centering
   \includegraphics[width=\textwidth]{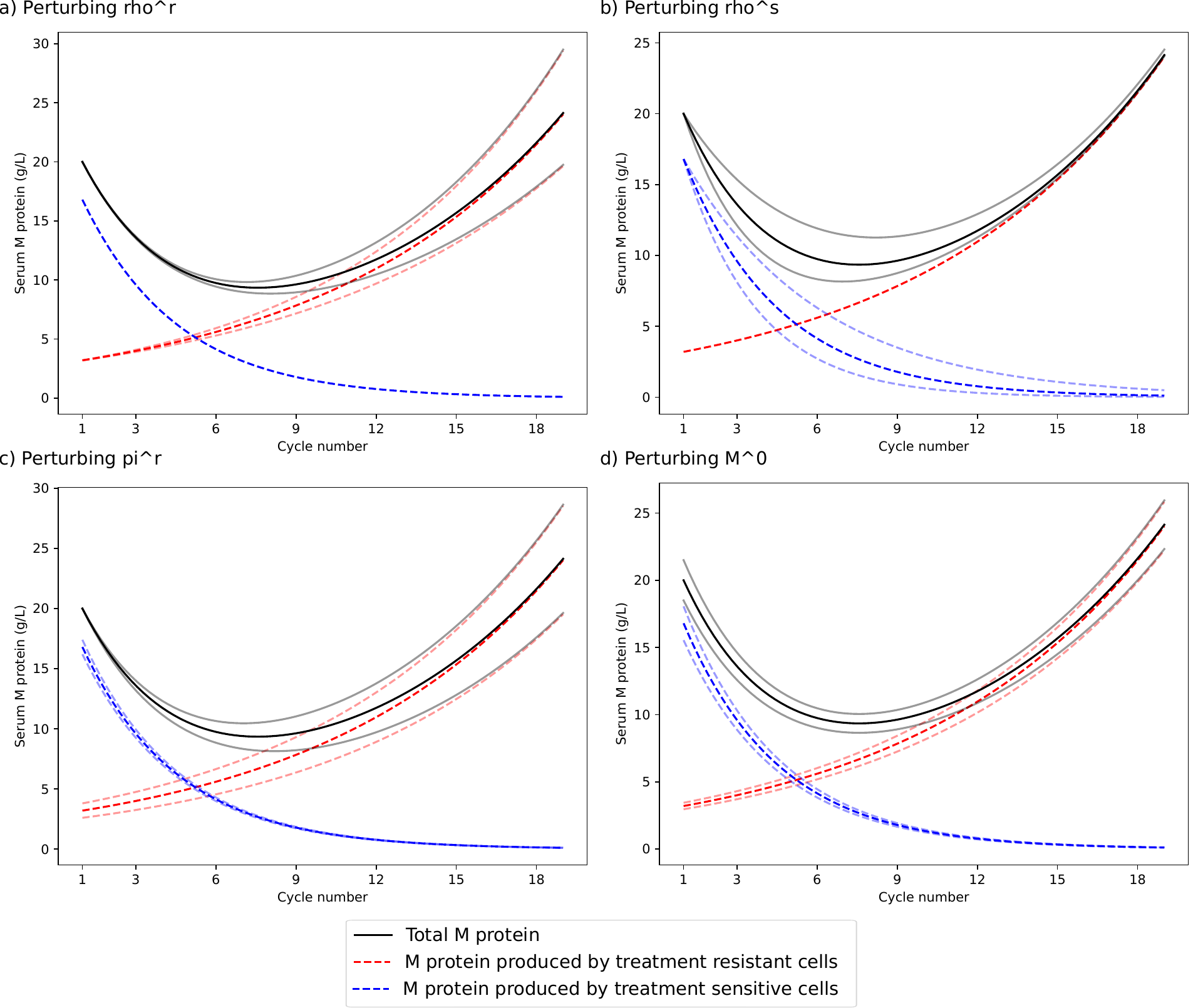}
   \caption{\textbf{Influence of model parameters on M protein trajectory.} 
   This figure shows the influence of small changes in the model parameters in equation \eqref{eq:model1} on the M protein trajectory of a patient with growth rate of the resistant population $\rho^r=0.004$, decay rate of the sensitive population $\rho^s=-0.01$, proportion of resistant cells at treatment start $\pi^r=0.16$, and M protein level at treatment start $M^0=20$. 
   \textbf{a)} $\rho^r$ is changed from $0.004$ to $0.0036$ and $0.0044$, respectively. 
   \textbf{b)} $\rho^s$ is changed from $-0.01$ to $-0.007$ and $-0.013$, respectively. 
   \textbf{c)} $\pi^r$ is changed from $0.16$ to $0.11$ and $0.21$, respectively. 
   \textbf{d)} $M^0$ is changed from $20$ to $18.5$ and $21.5$, respectively. 
   }
   \label{fig:example}
\end{figure}
\fi

In \cite{myklebust2024relapse}, a framework for prediction of relapse in MM patients from partially observed M protein trajectories was introduced and applied to a set of patients derived from the IKEMA trial \cite{moreau2021isatuximab}. 
The work introduced \modelnameplural{}, using equation \eqref{eq:model1} to model M protein observations, as well as relationships between baseline covariates and the mechanistic parameters $\pi^r$ and $\rho^r$. 
Interestingly, however, efforts to predict cancer dynamics under treatment using baseline covariates alone were unsuccessful, and inclusion of baseline covariates failed to improve the model accuracy. 

The models in \cite{myklebust2024relapse} used linear effects to model the relation between covariates and model parameters. 
Furthermore, the baseline covariates in \cite{myklebust2024relapse} were limited to blood tests, Fluorescence in situ hybridization (FISH), and demographic information like age and sex, which are known to hold little predictive value. 
More advanced sequencing techniques like gene expression profiling could contain the information needed to predict cancer dynamics under treatment, but would likely require modeling of nonlinear covariate effects or interactions between covariates. 

In this work, we present an extended version of the \modelname{} which uses a neural network to model the relation between covariates and model parameters, with the goal of capturing nonlinear covariate effects and interactions between covariates. 
Below, we describe the mentioned linear and non-linear covariate models.

\subsection{Joint modeling of patient-specific M protein trajectories}
Let $N$ be the number of patients, indexed by $i \in\left[1,\dots,N\right]$. 
Each patient has a vector of length $M_i$ of M protein measurements $y_{i} = (y_{i1}, \dots, y_{iM_i}) \in \R^{M_i}$. 
Each M protein observation $y_{ij}$ is measured at time $t_{ij}$, with $t_i = (t_{i1}=0, \dots, t_{iM_i}) \in \R^{M_i}$. 
Furthermore, denote the $p$ baseline covariates as $x_i = (x_{i1}, \dots, x_{ip}) \in \R^p$ for each patient.
M protein measurements $y_i$ are modeled as the sum of a resistant population growing exponentially with rate $\rho_i^r$, and a sensitive population decaying exponentially with rate $\rho_i^s$, initial total population size $M_i^0$, and initial proportion of resistant cells $\pi_i^r$, as shown in equation \eqref{eq:likelihood}. 
\begin{equation}
\begin{aligned}
    y_{ij} &=M(t_{ij}, \theta^i) + \varepsilon_{ij} \enspace , \enspace \varepsilon_{ij} \sim \mathcal{N} (0, \sigma^2) \\
    M(t_{ij}, \theta^i) &= M_i^0 \pi_i^r \exp\Big(\rho_i^r t_{ij}\Big) + M_i^0 (1-\pi_i^r) \exp \Big(\rho_i^s t_{ij}\Big) \\
\end{aligned}
\label{eq:likelihood}
\end{equation}
Note that $t_{i1} = 0$, which means that $y_{i1} = M(0, \theta_i) + \varepsilon_{i1} = M_i^0 + \varepsilon_{i1}$. 
Each patient has the following set of mechanistic parameters: 
\begin{itemize}
    \item $M_i^0$: True M-protein value at treatment start.
    \item $\pi_i^r$: Fraction of resistant cells at treatment start
    \item $\rho_i^r$: Growth rate of resistant cells.
    \item $\rho_i^s$: Growth rate of sensitive cells.
\end{itemize}
The mechanistic parameters are each physically required to be 
positive ($M^0$), 
between 0 and 1 ($\pi_i^r$), 
positive ($\rho_i^r$), and 
negative ($\rho_i^s$). 
To accommodate these physical constraints, the variables are transformed to corresponding transformed parameters with support on the entire real line. 
\begin{equation*}
\begin{aligned}
    \text{Mechanistic parameters } & \left[ \begin{array}{c} \rho_i^s \\ \rho_i^r \\ \pi_i^r \end{array} \right] = \left[ \begin{array}{c} -\exp(\theta_{1i}) \\ \exp(\theta_{2i}) \\ \frac{1}{1+e^{-\theta_{3i}}} \end{array} \right] \text{with domains } \begin{array}{c} \left[-\infty, 0\right] \\ \left[0,\infty\right] \\ \left[0,1\right] \end{array} .
\end{aligned}
\label{eq:model_parameters}
\end{equation*}
\begin{equation*}
\begin{aligned}
    \text{Model parameters } \theta^i &= \left[ \begin{array}{c} \theta_{1i} \\ \theta_{2i} \\ \theta_{3i} \end{array} \right] = 
    \left[ \begin{array}{c} \log(-\rho_{i}^s) \\ \log(\rho_i^r) \\ \log(\frac{\pi_i^r}{1-\pi_i^r}) \end{array} \right] \in \R^3. \\
\end{aligned}
\end{equation*}
By the model formulation in \eqref{eq:likelihood}, the observed $y_{i1}$ is normally distributed around $M_i^0$. 
M protein values are strictly non-negative. 
To use a prior with infinite support, a prior is placed on the log transformed $\theta_{4i} = \log(M_i^0)$ to make sure that all $M_i$ values are positive. 
\begin{equation}
\begin{aligned}
    \theta_{4i} &\sim \mathcal{N}\left(\log(y_{i1}), \xi^2\right) \\
    M_i^0 &= \exp(\theta_{4i}) \\
\end{aligned}
\end{equation}
The model uses a nonlinear mixed effect model (NLME) \cite{lindstrom1990nonlinear} framework to share information between patients, and sparsity-inducing priors to combat overfitting for covariate effects.
Using a nonlinear mixed effect framework (NLME) \cite{lindstrom1990nonlinear}, each patient has a random intercept which allows each patient to have different parameters $\rho_i^s, \rho_i^r$, and $\pi_i^r$. 
The patient-specific parameters $\theta_{1i}$, $\theta_{2i}$, and $\theta_{3i}$ are normally distributed around $\alpha_{1}$, $\alpha_{2}$, and $\alpha_{3}$ take a role similar to population averages. 
This choice of prior, which is described in equation \eqref{eq:nlme}, favors individual parameters that are similar within the group, while still allowing the model to fit to the data. The variation in each $\theta_l$ within the population is parametrized by the standard deviation hyperparameter $\omega_l$ for $l \in \{1,2,3\}$. 
\begin{equation}
\begin{aligned}
    \text{For } l=1,\dots,3: \enspace \theta_{li} \sim \mathcal{N}(\alpha_l, \omega_l^2) \\
\end{aligned}
\label{eq:nlme}
\end{equation}
To set priors for the parameters $\alpha_l$, $l=1,2,3$, a guess of each average $\theta_{li}$ in the population is required. The prior is then defined as in equation \eqref{eq:alphaprior}.
\begin{equation}
\begin{aligned}
    \alpha_l \sim \mathcal{N}(\widehat{\alpha_l} , 1) \enspace , \enspace l \in [1,2,3]
\label{eq:alphaprior}
\end{aligned}
\end{equation}

For $\sigma$ and $\omega$, the following priors are used: 
\begin{equation}
\begin{aligned}
    \sigma &\sim \text{HalfNormal}(0,1)  \enspace : \enspace \sigma > 0 \\
    \xi &\sim \text{HalfNormal}(0,1)  \enspace : \enspace \xi > 0 \\
    \omega_l &\sim \text{HalfNormal}(0,1)  \enspace : \enspace \omega_l > 0 \enspace, \enspace l \in [1,2,3]
\end{aligned}
\end{equation}
\subsection{Linear covariate effects}
To model relationships between baseline covariates and mechanistic parameters, a function that maps baseline covariates to an effect on each mechanistic parameter can be included in the model, as in equation \eqref{eq:covariate_mapping}, where $x$ is the covariate vector with dimension $p$: 
\begin{equation}
\begin{aligned}
    \text{For } l \in \{1,2,3\}: \enspace \theta_{li} \sim \mathcal{N}(\alpha_l + f_l(x), \omega_l^2), \enspace \beta_l = (\beta_{l1}, \dots, \beta_{lp}) \in \R^p \\
\end{aligned}
\label{eq:covariate_mapping}
\end{equation}
In the linear covariate model, the effect of the $p$ covariates in $x$ on parameter $\theta_l$ is modeled by the function $f_l(x) = x\beta_l^T$, and the full expression becomes: 
\begin{equation}
\begin{aligned}
    \text{For } l \in \{1,2,3\}: \enspace \theta_{li} \sim \mathcal{N}(\alpha_l + x\beta_l^T, \omega_l^2), \enspace \beta_l = (\beta_{l1}, \dots, \beta_{lp}) \in \R^p \\
\end{aligned}
\label{eq:nlme_linear_covariate_effect_model}
\end{equation}
To accommodate automatic discovery of covariate effects, a sparsity inducing prior is used for the coefficients $\beta_l$, namely the hierarchical regularized horseshoe \cite{piironen2017sparsity}. 
The hierarchical regularized horseshoe introduces a global shrinkage parameter $\tau_l$ and a set of local shrinkage parameters $\tilde{\lambda}_{lb}$; one for each $\beta_{lb}, b \in [1,\dots,p]$. 
It also requires an initial guess of the number of nonzero parameters, $p_0$. 
This is chosen to be $\widehat{p_0} = \text{int}(p/2)$. 
For a more efficient sampling, we reparametrize $\beta_{lb}$ as $\beta_{lb} = z_{lb} \tau_l \tilde{\lambda}_{lb}$. 
This means that $z_{lb} \sim \mathcal{N}(0,1)$. 
The following priors are used: 
\begin{equation}
\begin{aligned}
    \beta_{lb} &\sim \mathcal{N}(0,\tau_l^2 \cdot \Tilde{\lambda}_{lb}^2) \enspace , \enspace l \in [1,2,3] \enspace , \enspace b \in [1,\dots,p] \\
    \tau_l &\sim \text{Half-StudentT}_2\left(\frac{p_0}{p-p_0}\cdot \frac{\sigma}{\sqrt{N}}\right) \enspace , \enspace l \in [1,2,3] \\
    \Tilde{\lambda}_{lb}^2 &= \frac{c^2\lambda_{lb^2}}{c^2 + \tau_l^2\lambda_{lb}^2}  \\
    \lambda_{lb} &\sim \text{Half-StudentT}_2(1) \enspace , \enspace l \in [1,2,3] \enspace , \enspace b \in [1,\dots,p] \\
    c^2 &\sim \text{InverseGamma}(1,1) \\
\end{aligned}
\label{eq:priors}   
\end{equation}

\subsection{Interactions and nonlinear covariate effects}
Modeling the effects of the covariates using a linear model makes inference easy, but the linear model is not able to capture nonlinear effects or interaction effects between variables. 
Given $p$ covariates, interaction effect can be naively modeled by taking advantage of the sparsity enforcing prior by adding the $p(p-1)$ possible interactions to the covariate vector. 
However, the number of covariates scales poorly with $p$, and complicates the Bayesian inference. 
Furthermore, such an approach would only model linear interaction effects. 

To model nonlinear effects of covariates and interactions between covariates, we will introduce a \modelname{} with a shallow neural network mapping from covariates to model parameters. 
With this change, equation \eqref{eq:covariate_mapping} becomes: 
\begin{equation}
\begin{aligned}
    \text{For } l=1,\dots,3: \enspace \theta_{li} \sim \mathcal{N}(\alpha_l + f_l(\vect{x}_i|\vect{z}_l,L,K,g), \omega_l^2)
\end{aligned}
\end{equation}
where $f_l(\vect{x}_i|\vect{z}_l,L,K,g)$ is the output of a fully connected feedforward neural network with $L$ layers, $K$ nodes per layer, weights $\vect{z}_l$ and activation function $g(\cdot)$. 
Inspired by \cite{jylanki2014expectation}, we choose a 2-layer neural network ($L=2$) with two hidden nodes ($K=2$), the smallest architecture capable of capturing covariate interactions. 
With this model choice, we have: 
\begin{equation}
\begin{aligned}
    f_l(\vect{x}_i|\vect{z}_l,L=2,K=2,g) = \sum_{k=1}^2 v_k g(\vect{w}_k^T \vect{x}_i) + v_0
\end{aligned}
\label{eq:node}
\end{equation}
where $k$ is the index of the hidden node, $\vect{w}_k = [w_{k,1}, \dots, w_{k,p}]^T$ are the weights from the $p$ covariates to the hidden nodes, $v_k$ is the weight from the hidden node to the output layer, and $\vect{z}_l = [\vect{w}^T, \vect{v}^T]^T$, where $\vect{w} = [\vect{w}_1^T, \dots, \vect{w}_k^T]^T$ and $\vect{v} = [v_1, \dots, v_k, v_0]^T$.
The leaky rectified linear unit (Leaky ReLU) function \cite{maas2013rectifier} is used as activation function: 
\begin{equation}
\begin{aligned}
    g(x) = 
    \begin{cases} 
    x, & \text{if } x > 0 \\
    0.1 x, & \text{otherwise}
    \end{cases} \\
\end{aligned}
\end{equation}
To select priors for the neural network weights, the naive approach is to give each element in $\vect{z}_l$ an isotropic prior: 
\begin{equation}
\begin{aligned}
    \vect{z}_l \sim \mathcal{N}(\vect{0};\sigma_{z}\matr{I}) &\qquad \text{with} \qquad \sigma_{z} \sim \text{HalfNormal}(0,1)
\end{aligned}
\end{equation}
where $\vect{0}$ is a zero vector of the same dimension as $\vect{z}_l$.
Isotropic Gaussians are the most commonly used priors for the weights in neural networks, but introduce non-identifiability issues even in architectures as small as two layers. 
To combat these non-identifiability issues, we constrain the values of the output weights to be positive by using Half Student T priors with $\nu = 2$ degrees of freedom and $\sigma = 1$, which has the following probability density function for $x \geq 0$:
\begin{equation}
    f(x; \nu) = \frac{2\Gamma\left(\frac{\nu + 1}{2}\right)}{\Gamma\left(\frac{\nu}{2}\right)\sqrt{\nu\pi\sigma^2}}\left(1 + \frac{x^2}{\nu \sigma^2}\right)^{-\frac{\nu + 1}{2}}
\end{equation}
where $\Gamma$ denotes the gamma function.

\subsection{Simulation study}
This section describes the design of a simulation study with the aim to compare two \modelnameplural{}, one with a linear mapping from covariates to model parameters, and one with a BNN mapping. 
The mechanistic model parameters of the simulated patients will be generated with a system that contains interaction effects.
Both models will be trained on M protein measurements and baseline covariates of \Npatients{} training patients, then evaluated by their ability to predict M protein trajectories for \Ntestpatients{} test patients using only baseline covariates and the M protein value at treatment start. 

\Npatients{} patients were simulated, with M protein measured every 28 days for a maximum of \Nmeasurements{} cycles, or \Ndays{} days.
To more closely resemble real data, random loss to follow-up was introduced by sampling a random date off loss to follow-up. 
The number of measurements observed for a patient before loss to follow-up was sampled randomly from a uniform distribution from 5 to \Nmeasurements{} measurements. 
Each patient will have five covariates measured at baseline, named covariate 1, covariate 2, covariate 3, covariate 4 and covariate 5, respectively. 
What each covariate represents is undefined by intent, as the nature of each covariate is unimportant for the model comparison. 
For each patient, the values of each covariate was sampled from an uniform distribution, as follows: 
\begin{equation}
\begin{aligned}
        x_{i,j} \sim Unif([-1,1]), j \in {1, \dots, 5}  \\
\end{aligned}
\label{eq:unif_x}
\end{equation}
Out of the five covariates, only the first three had an effect on model parameters $\rho^r$ and $\pi^r$, as shown in Table \ref{tab:cov_effects_baseline_study}. 
Importantly, for $\rho^r$, there is an interaction effect between covariates 1 and 2, and 
for $\pi^r$, there is an interaction effect between covariates 2 and 3. 
Covariates 4 and 5 do not have any effect, and are included to test the feature selecting capacities of the models. 
\begin{table}[H]
    \caption{Covariate effects} \label{tab:cov_effects_baseline_study} 
    \centering
    \begin{tabular}{|c|c|c|}
    \hline
        Effects on $\rho^r$ & & \\ \hline 
        Covariate or interaction & Parameter & Value  \\ \hline
        Covariate 1 & $\beta_{\rho,1}$ & 0.4 \\ \hline
        Covariate 2 & $\beta_{\rho,2}$ & 0 \\ \hline
        Covariate 3 & $\beta_{\rho,3}$ & 0 \\ \hline
        Covariate 4 & $\beta_{\rho,4}$ & 0 \\ \hline
        Covariate 5 & $\beta_{\rho,5}$ & 0 \\ \hline
        Interaction between 1 and 2 & $\beta_{\rho}^*$ & -0.8 \\ \hline
    \hline
        Effects on $\pi^r$ & & \\ \hline
        Covariate or interaction & Parameter & Value \\ \hline
        Covariate 1 & $\beta_{\pi,1}$ & 0 \\ \hline
        Covariate 2 & $\beta_{\pi,2}$ & 0.4 \\ \hline
        Covariate 3 & $\beta_{\pi,3}$ & -0.6 \\ \hline
        Covariate 4 & $\beta_{\pi,4}$ & 0 \\ \hline
        Covariate 5 & $\beta_{\pi,5}$ & 0 \\ \hline
        Interaction between 1 and 2 & $\beta_{\pi}^*$ & 1\\ \hline
    \end{tabular} \\
\end{table}
\noindent Using the covariate value of each patient, and the covariate effects in Table \ref{tab:cov_effects_baseline_study}, the values of $\rho_i^r$ and $\pi_i^r$ for each patient were determined using equation \eqref{eq:generate_model_parameters}. The value of $\rho_i^s$ for each patient was set to $-0.04$. 
\begin{equation}
\begin{aligned}
    \theta_{\rho_r}^i &=\log(0.002) + x_i\beta_{\rho}^{T} + x_{i,1}x_{i,2}\beta_{\rho}^{*}, \enspace \beta_{\rho} = (\beta_{\rho,1}, \dots, \beta_{\rho,5}) \in \R^5 \\
    \rho_i^r &= \exp(\theta_{\rho_r}^i) \\
    \theta_{\pi}^i &=\log(0.1) + x_i\beta_{\pi}^{T} + x_{i,2}x_{i,3}\beta_{\rho}^{*}, \enspace \beta_{\pi} = (\beta_{\pi,1}, \dots, \beta_{\pi,5}) \in \R^5 \\
    \pi_i^r &= \exp(\theta_{\pi}^i) / (\exp(\theta_{\pi}^i) + 1) \\
\end{aligned}
\label{eq:generate_model_parameters}
\end{equation}
Figure \ref{fig:covariate_effects} shows how the true model parameters $\rho^r$ and $\pi^r$ in the population depend on the values of covariates 1, 2 and 3. 
For $\rho^r$, the interaction effect between covariates 1 and 2 is shown by the fact that the marginal effect of covariate 2 changes with the value of covariate 1: 
Among patients with \textit{lower} values in covariate 1, the $\rho^r$ value \textit{increases} with the value of covariate 2;
conversely, 
among patients with \textit{higher} values in covariate 1, the $\rho^r$ value \textit{decreases} with the value of covariate 2;
For $\pi^r$, there is also an interaction effect, but between covariates 2 and 3. 
\ifshowfigures
\begin{figure}[htbp]
   \centering
   \includegraphics[width=\textwidth]{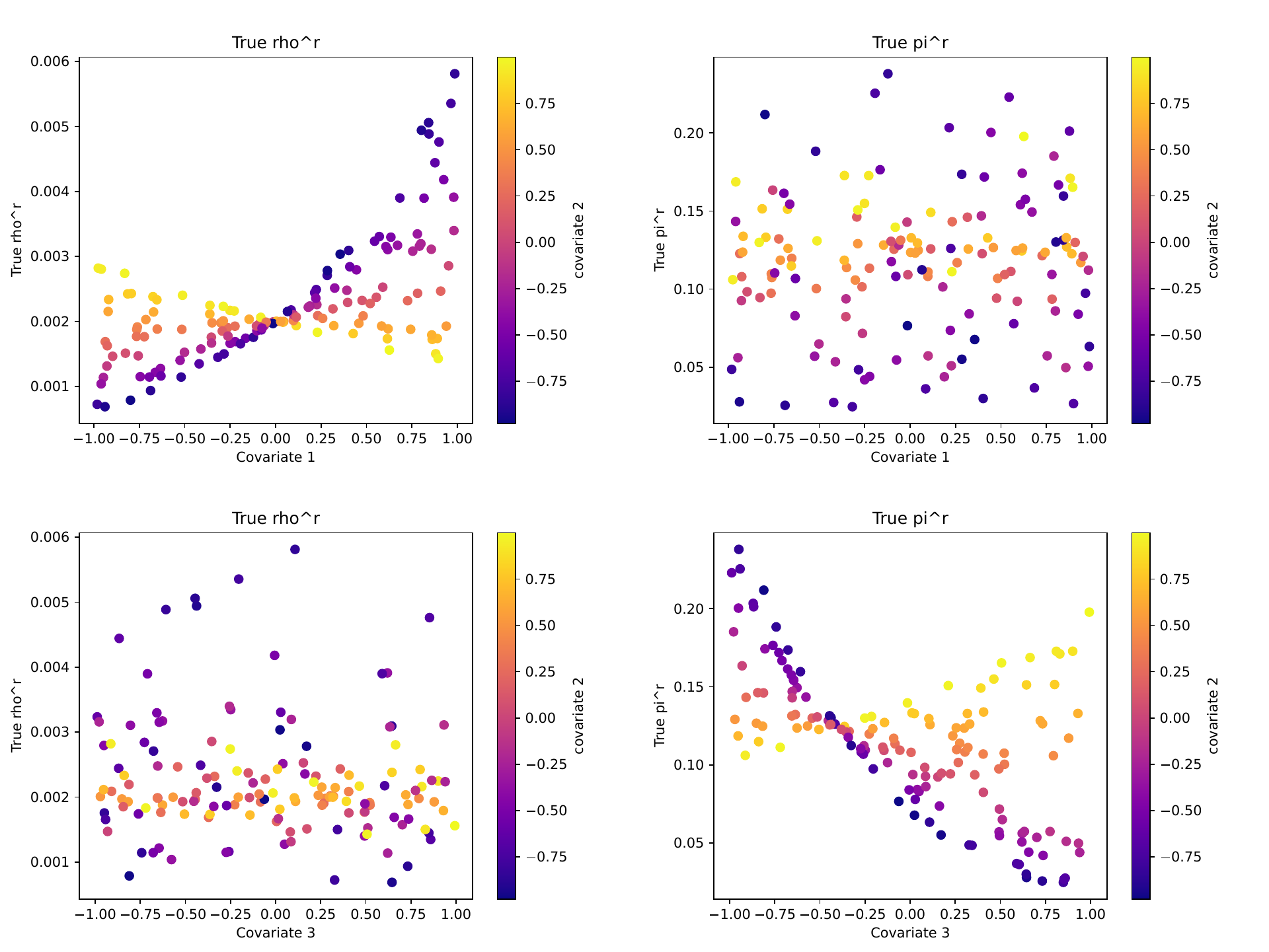}
   \caption{Effects of covariates 1, 2, and 3 on model parameters $\rho^r$ and $\pi^r$. 
   These scatter plots show values of $\rho^r$ (left) and $\pi^r$ (right) for all patients. 
   In the top row, the x axis represent covariate 1; 
   in the bottom row, the x axis represent covariate 3. 
   The dots are colored according to the value of covariate 2.
   }
   \label{fig:covariate_effects}
\end{figure}
\fi

For each patient, the initial M protein value and the values of the five covariates were sampled randomly according to the distributions in equation \eqref{eq:dist_mprot_start}. 
To simulate measurement error, normally distributed, independent and zero centered noise terms with standard deviation \noiselevel{} were sampled randomly and added to each simulated M protein measurement. 
\begin{equation}
\begin{aligned}
        \log(M_i^0) \sim \mathcal{N}(\log(50), 0.1)   \\
\end{aligned}
\label{eq:dist_mprot_start}
\end{equation}
To make the inference more challenging, a natural variation in model parameters can be introduced by adding i.i.id normally distributed random effects to the model parameters of each patient using the equations in \eqref{eq:randomeffects}. 
The normally distributed random effects can be zero-centered and have variance $\omega_l$, for $l \in \{1,2,3\}$. 
\begin{equation}
\begin{aligned}
    \theta_{\rho_s}^i &\sim \mathcal{N}(\log(0.04), \omega_1^2) \\
    \rho_i^s &= -\exp(\theta_{\rho_s}^i) \\
    \theta_{\rho_r}^i &\sim \mathcal{N}(\log(0.002) + x_i\beta_{\rho}^{T} + x_{i,1}x_{i,2}\beta_{\rho}^{*}, \omega_2^2), \enspace \beta_l = (\beta_{\rho,1}, \dots, \beta_{\rho,p}) \in \R^p \\
    \rho_i^r &= \exp(\theta_{\rho_r}^i) \\
    \theta_{\pi}^i &\sim \mathcal{N}(\log(0.1) + x_i\beta_{\pi}^{T} + x_{i,2}x_{i,3}\beta_{\rho}^{*}, \omega_3^2), \enspace \beta_l = (\beta_{\pi,1}, \dots, \beta_{\pi,p}) \in \R^p \\
    \pi_i^r &= \exp(\theta_{\pi}^i) / (\exp(\theta_{\pi}^i) + 1) \\
\end{aligned}
\label{eq:randomeffects}
\end{equation}

\subsection{Inference and evaluation}
To fit the BNN and linear covariate effect models to the data, 10,000 Markov Chain Monte Carlo (MCMC) samples were sampled from the posterior of each model using Hamiltonian Monte Carlo (HMC). 
All statistical analysis was performed in Python version 3.11.3.
Posterior sampling was performed in PyMC version 5.1.1 \cite{salvatier2016probabilistic} using the No-U-Turn sampler (NUTS) introduced in \cite{hoffman2014no}. 
To improve convergence, the sampler was initialized using automatic differentiation variational inference (ADVI) \cite{kucukelbir2017automatic} within the PyMC software. 
The code for this project is available at \href{https://github.com/evenmm/mm-predict-bnn}{https://github.com/evenmm/mm-predict-bnn}.

\section{Results}
Figure \ref{fig:all_mpro_cas1} shows the simulated M protein trajectories of the entire population in the simulation study without random effects in model parameters and noise standard deviation of 1. 
Figure \ref{fig:train_and_test} shows examples of how the \modelnameplural{} with linear and nonlinear covariate effects fit training data and their predictive performance on test data. 
In Figure \ref{fig:median_error_boxplots}, the absolute errors for the two models are compared for various cycles after treatment start, showing that the BNN method also outperforms the linear model numerically. 

For the simulation study in figures \ref{fig:all_mpro_cas1}, \ref{fig:train_and_test}, and \ref{fig:median_error_boxplots}, the noise was normally distributed with a standard deviation of 1, which is a feasible noise level for M protein measurements, but the methods could perform differently at different noise levels. 
To test how the BNN and linear covariate models performed under different noise levels, the simulation study was repeated with a noise standard deviation of 2. 
Furthermore, if the model parameters $\rho^r$, and $\pi^r$ were not deterministically determined by the values of the covariates, but have random individual variations as described in equation \ref{eq:randomeffects}, this would be even more challenging. 

To investigate how well the model would work at higher noise levels and with random effects in individual model parameters, additional simulation studies were carried out, adding random variations to individual model parameters, according to equation \ref{eq:randomeffects}. 
The random effect variances $\omega_l$ were set to 0.10, 0.05, and 0.20 for $\rho^s$, $\rho^r$ and $\pi^r$, respectively. 
The simulated data under the different experimental settings are shown in figure \ref{fig:all_mpro}. 
Model fits to example training patients and predictions on test patients corresponding to figure \ref{fig:train_and_test} for alternative experimental settings are provided in the supplementary material. 
Figure \ref{fig:all_boxes} shows the predictive performance of the linear and BNN covariate effect models for these experimental settings. 

\ifshowfigures
\begin{figure}[htbp]  
   \centering
   \includegraphics[width=\textwidth]{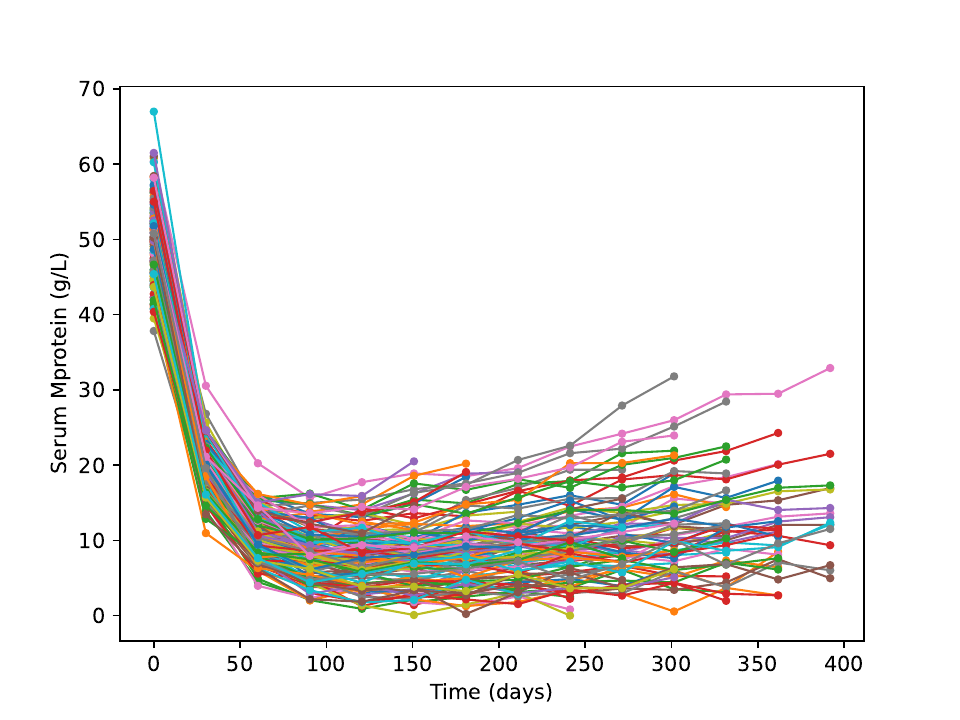}
   \caption{\textbf{M protein observations in the simulated training population.} 
   All simulated M protein observations for the \Npatients{} patients in the training set are shown together, colored per patient.}
   \label{fig:all_mpro_cas1}
\end{figure}
\begin{figure}[htbp]
   \centering
   \includegraphics[width=\textwidth]{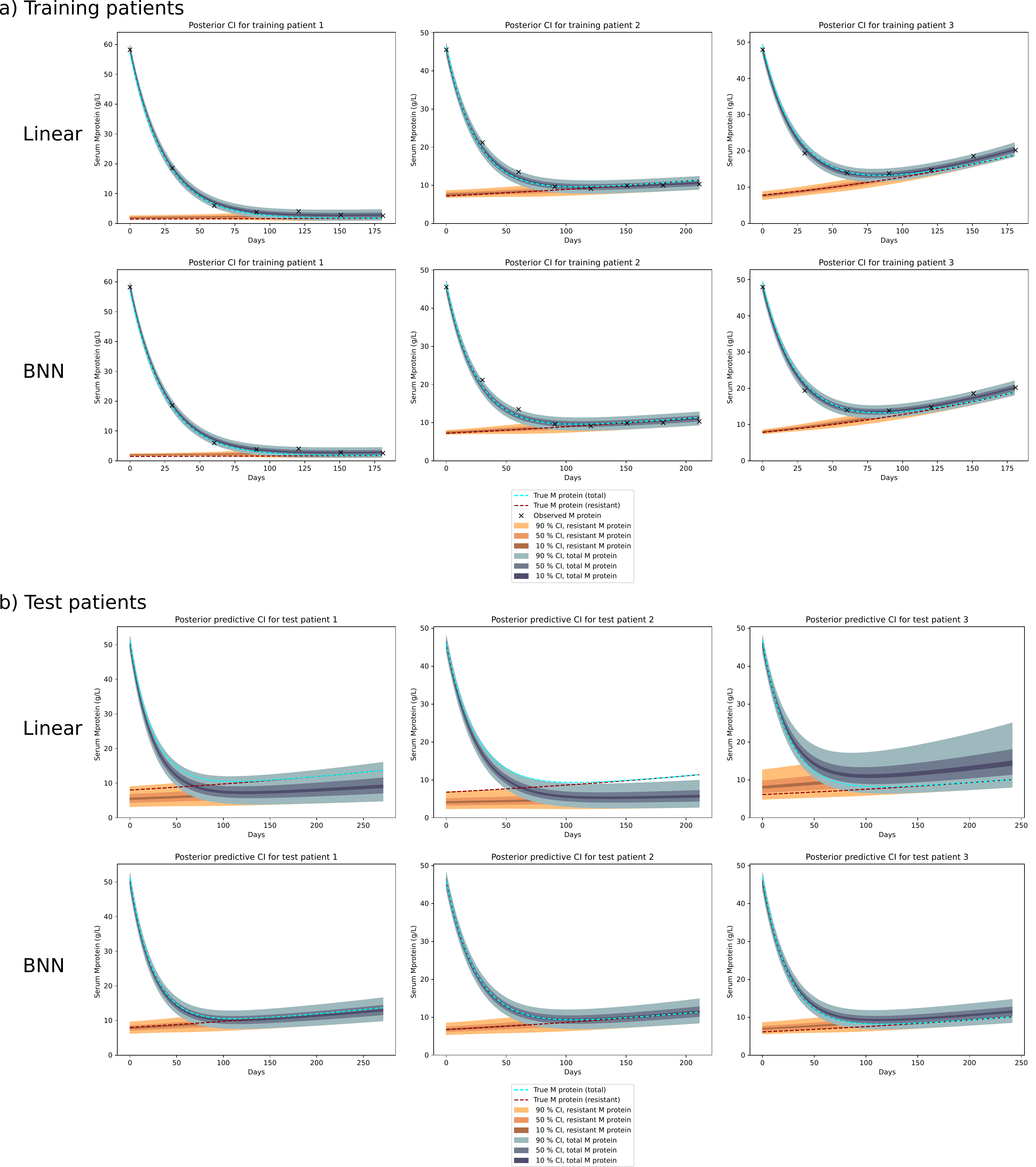}
   \caption{\textbf{Model fit to training data and predictions on test data.} a) Model fit of the linear and the BNN covariate effect models to patients in the training set, where M protein measurements and baseline covariates were provided to the models.
   b) Predictions on test patients, where only baseline covariates and initial M protein were provided, and the entire M protein trajectory was predicted by each model. 
   The standard deviation of the observation noise is equal to 1, and there are no random effects in the model parameters of patients in the training or the test set.}
   \label{fig:train_and_test}
\end{figure}
\begin{figure}[htbp]  
   \centering
   \includegraphics[width=\textwidth]{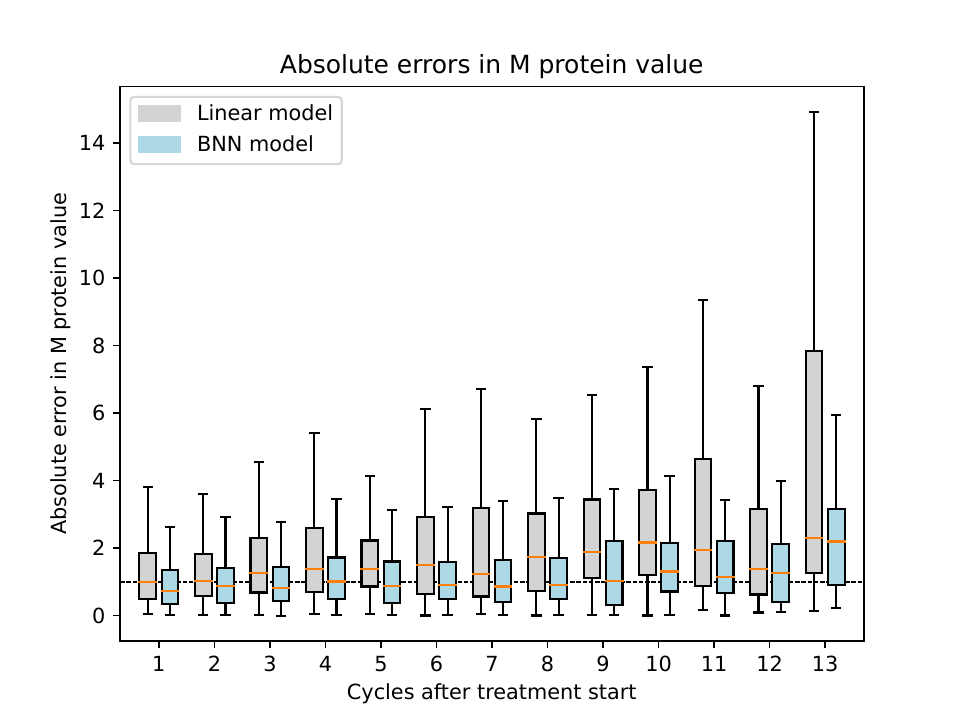}
   \caption{\textbf{Comparison of predictive performance between the linear and BNN covariate effect models.} Box plots of the absolute difference between M protein observations and posterior median of predicted M protein value at every 28 day cycle since treatment start, for the \Ntestpatients{} test patients. The dotted line marks the standard deviation of the observation noise.}
   \label{fig:median_error_boxplots}
\end{figure}
\begin{figure}[htbp]  
   \centering
   \includegraphics[width=\textwidth]{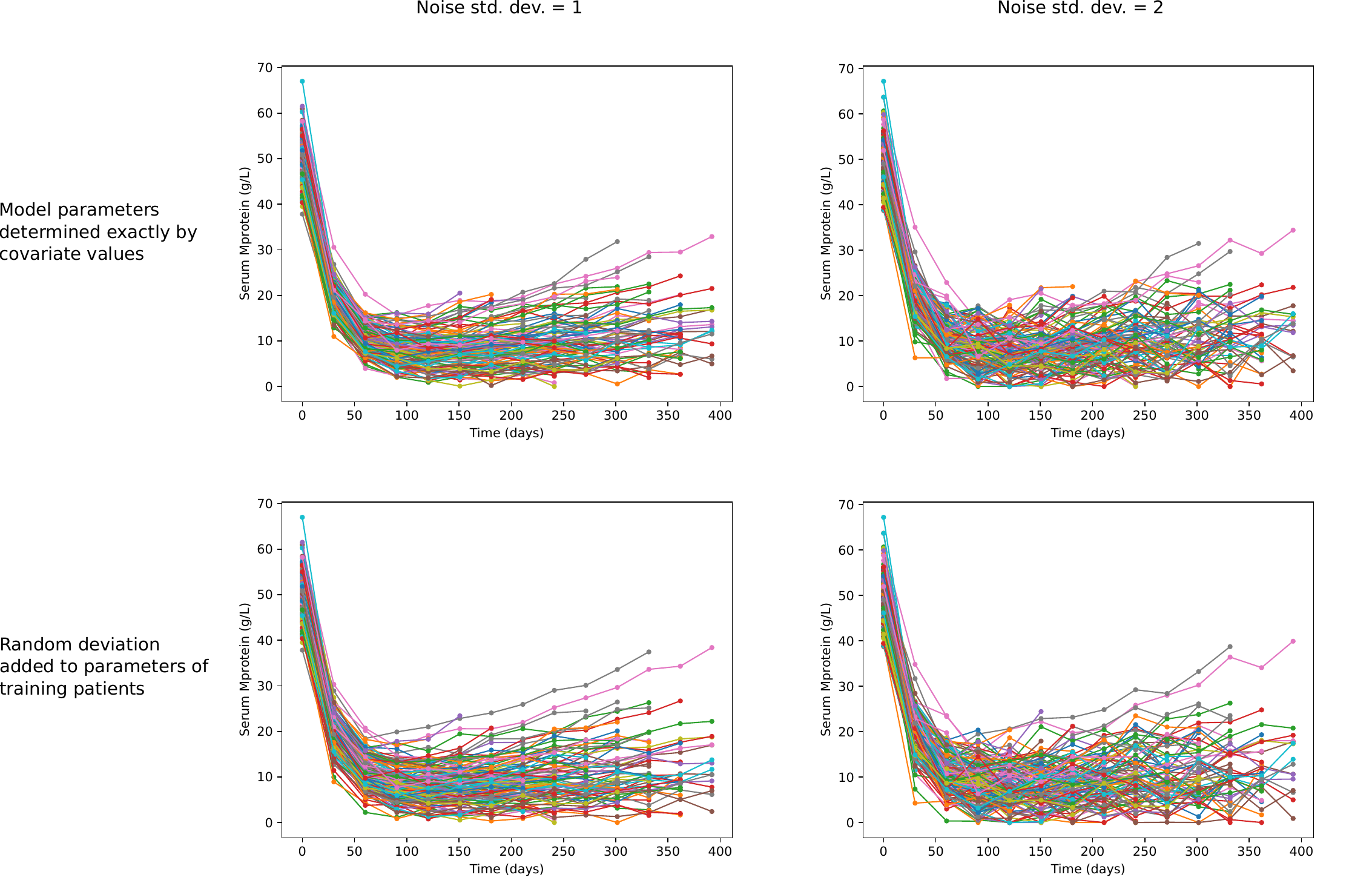}
   \caption{\textbf{Simulated training population under different experimental settings.} 
   Each subpanel shows all simulated M protein observations for all \Npatients{} patients in the training set, with and without random effects in the model parameters, and at different noise levels.}
   \label{fig:all_mpro}
\end{figure}
\begin{figure}[htbp]  
   \centering
   \includegraphics[width=\textwidth]{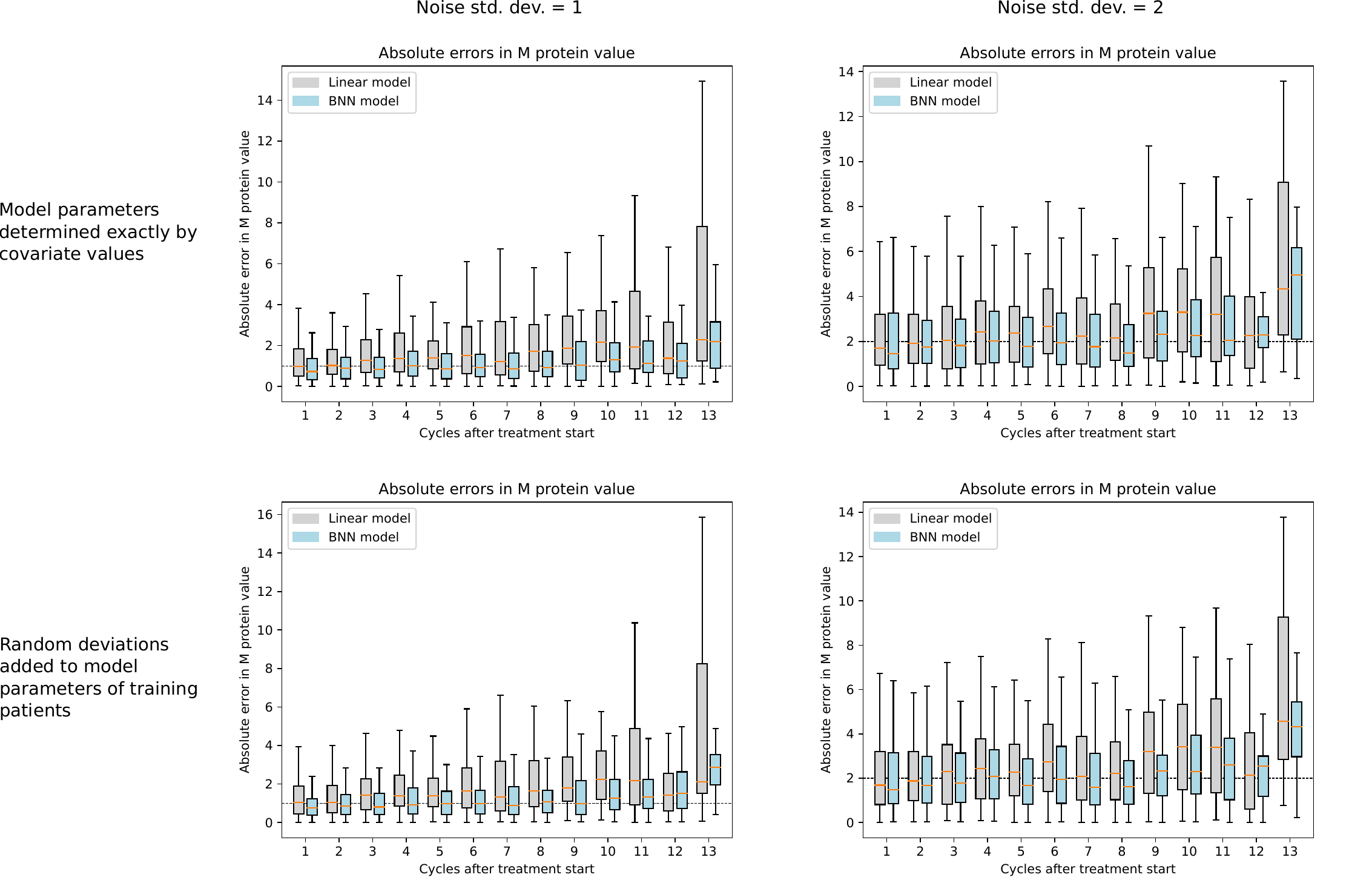}
   \caption{\textbf{Comparison of predictive performance under different experimental settings.} 
   Each subpanel shows box plots of the absolute difference between M protein observations and posterior median of predicted M protein value for the \Ntestpatients{} test patients, with and without random effects in the model parameters of training patients, and at different noise levels. In each subpanel, the dotted line marks the standard deviation of the observation noise in the setting.}
   \label{fig:all_boxes}
\end{figure}
\fi

\section{Discussion}
The purpose of the simulation study was to evaluate whether the BNN model is able to predict M protein trajectories using only baseline covariates and whether it is able to outperform a linear model in the presence of covariate interaction effects. 
Figure \ref{fig:train_and_test} shows that both the BNN and the linear covariate effect models fit well with the observed data for training patients and that the BNN model provides better predictions for test patients. 
In figure \ref{fig:all_boxes}, the standard deviation of the noise level is included in the interquartile range of errors of the BNN covariate model, for all cycles after starting treatment except 13, which is the longest time frame. 
This shows that the errors are comparable to the inherent noise in the system. 

The baseline covariates presented in this study are intentionally unspecified and unnamed to underline the modular structure of the theoretical framework. 
The covariate effects in the simulation study are very strong. 
In real-world data, the size of the random effects relative to the covariate effects could be larger than in the simulated data.
Random effects in the model parameters were only introduced for training patients. 
The introduction of random effects for test patients would almost surely lead to a reduction in predictive performance on test data for both models. 
However, to compare the two methods against each other, it is more useful not to add random variations to the test patients, since these could anyway not be predicted by either of the models. 

As an example, we have applied our method to multiple myeloma (MM).  
In MM, predictive and prognostic biomarkers remain elusive. 
One of the advances in the treatment of multiple myeloma during the last decade is CD38 inhibitors such as daratumumab \citep{lokhorst2015targeting} and isatuximab \citep{frampton2021isatuximab}. 
CD38 inhibitors are typically used in combination with Dexamethasone and either Pomalidomide or Carfilzomib. 
Within these combinations, there could be individual differences in how well patients respond to the proteasome inhibitors Pomalidomide and Carfilzomib. 
When informative covariates are available, the models presented in this study could be used to predict the dynamics of cancer under a variety of treatments, thus helping in the choice of treatment for each patient.
As an example, they could inform on the choice of proteasome inhibitor to use in combination with the CD38 inhibitor dexamethasone for the treatment of multiple myeloma. 

In \cite{myklebust2024relapse}, future M protein values were predicted using previously observed M protein values and covariate effects.
If provided with such informative covariates that the entire M protein trajectory from the beginning of a treatment could be predicted using covariates only, the methods developed in the current work would not only improve predictions when M protein values are partially observed, but could allow prediction of M protein trajectories from baseline. 
With this comes the opportunity of using the model as a tool for treatment selection, by comparing trajectories under potential treatments. 

We have developed a novel \modelname{} for the prediction of cancer dynamics under treatment. 
Our results show that the BNN covariate effect model is superior to the linear covariate effect model for prediction in the presence of covariate interaction effects, even at higher noise levels and with model parameters that are not directly determined by the baseline covariates. 
In summary, the BNN model captures interactions between covariates while maintaining uncertainty quantification, accurately predicting M protein trajectories using only simulated covariates. 
This shows the promise of \modelnameplural{} for treatment suggestion when provided with strongly predictive covariates.

\section*{Acknowledgments}
Simulation studies were performed on resources provided by UNINETT Sigma2 - the National Infrastructure for High-Performance Computing and Data Storage in Norway, and by Oslo Centre for Biostatistics and Epidemiology. 
The study has benefited from data provided by Oslo Myeloma Center and the COMMPASS project to sharpen questions and develop models. 
The authors thank Anna Lysén at Oslo Myeloma Center for her assistance in providing access to data which inspired the development of the models presented in this work. 
E. M. Myklebust, A. Köhn-Luque, and A. Frigessi were supported by the Center for research-based-innovation BigInsight under grant 237718 from the Research Council of Norway. 
J. Foo and K. Leder were supported by the Fulbright US-Norway Foundation. 
J. Foo and K. Leder were supported by the University of Oslo-University of Minnesota Norwegian Centennial Chair Grant. 
J. Foo was supported by the US National Science Foundation under grant number DMS-2052465. 
K. Leder was supported by the US National Science Foundation under grant number CMMI-1552764. 
This work was also supported by the Research Council of Norway through the projects DL: Pipeline for individually tailoring new treatments in hematological cancers (PINpOINT) under project number 294916; INTPART-International Partnerships for Excellent Education and Research under project number 309273; and Integreat - Norwegian Centre for knowledge-driven machine learning under project number 332645.
The authors also acknowledge the Centre for Digital Life Norway for supporting the partner project PINpOINT.

\section*{CRediT author contribution statement}
\textbf{E. M. Myklebust:} Conceptualization, Methodology, Software, Validation, Formal analysis, Investigation, Data curation, Writing – original draft, Writing – review $\&$ editing, Visualization.
\textbf{A. Frigessi:} Conceptualization, Methodology, Writing – review $\&$ editing, Supervision, Funding acquisition.
\textbf{F. Schjesvold:} Conceptualization, Methodology, Writing – review $\&$ editing.
\textbf{J. Foo:} Conceptualization, Methodology, Writing – review $\&$ editing, Supervision.
\textbf{K. Leder:} Conceptualization, Methodology, Writing – review $\&$ editing, Supervision.
\textbf{A. Köhn-Luque:} Conceptualization, Methodology, Writing – original draft, Writing – review $\&$ editing, Supervision.

\subsection*{Declaration of Interests}
F.S. received honorarium from Sanofi, Janssen, BMS, Oncopeptides, Abbvie, GSK, and Pfizer.
The authors declare that they have no other conflicts of interest.

\clearpage
\bibliographystyle{unsrtnat}
\bibliography{m2p}

\clearpage
\section*{Supplementary material}

\ifshowfigures
\begin{figure}[htbp]
   \centering
   \includegraphics[width=0.88\textwidth]{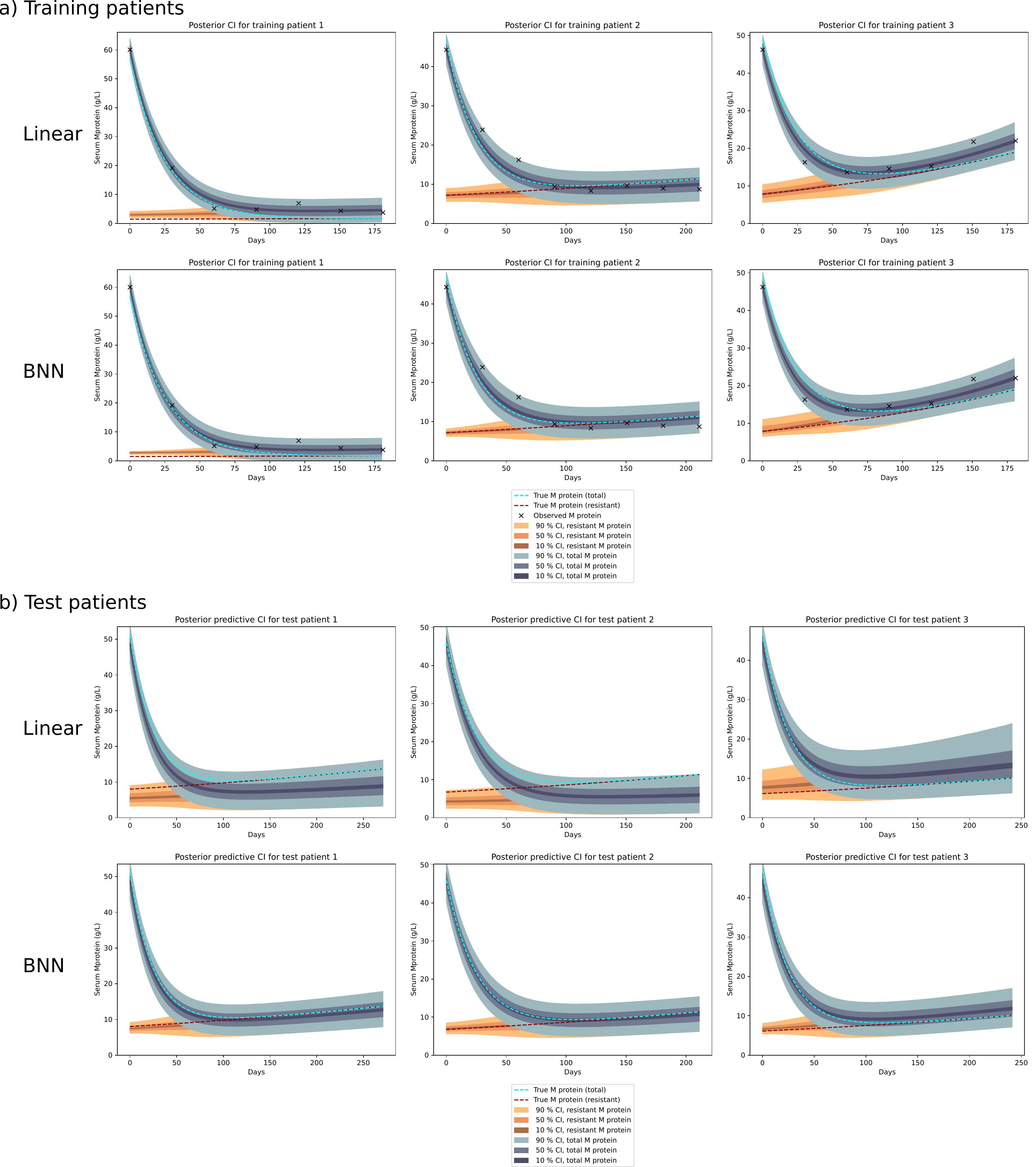} 
   \caption{\textbf{Model fit to training data and predictions on test data, with noise standard deviation = 2 and no random effects in model parameters of training patients.} a) Model fit of the linear and BNN covariate effect models to patients in the training set, where M protein measurements and baseline covariates were provided to the models.
   b) Predictions on test patients, where only the baseline covariates were provided to the models.}
   \label{fig:train_and_test_2}
\end{figure}
\fi

\ifshowfigures
\begin{figure}[htbp]
   \centering
   \includegraphics[width=\textwidth]{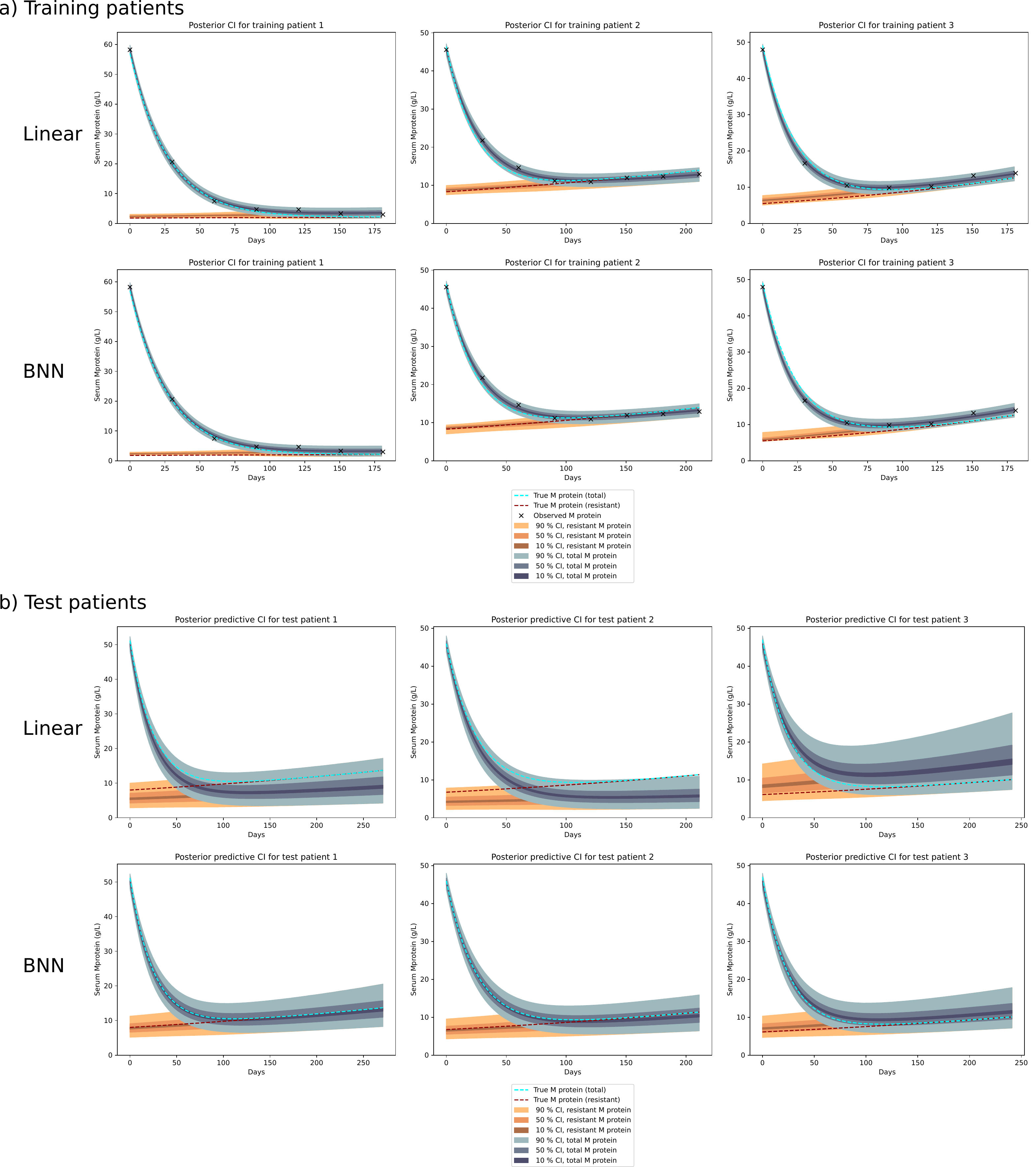}
   \caption{\textbf{Model fit to training data and predictions on test data, with noise standard deviation = 1 and \textit{with} random effects in model parameters of training patients.} a) Model fit of the linear and BNN covariate effect models to patients in the training set, where M protein measurements and baseline covariates were provided to the models.
   b) Predictions on test patients, where only the baseline covariates were provided to the models.}
   \label{fig:train_and_test_4}
\end{figure}
\fi

\ifshowfigures
\begin{figure}[htbp]
   \centering
   \includegraphics[width=\textwidth]{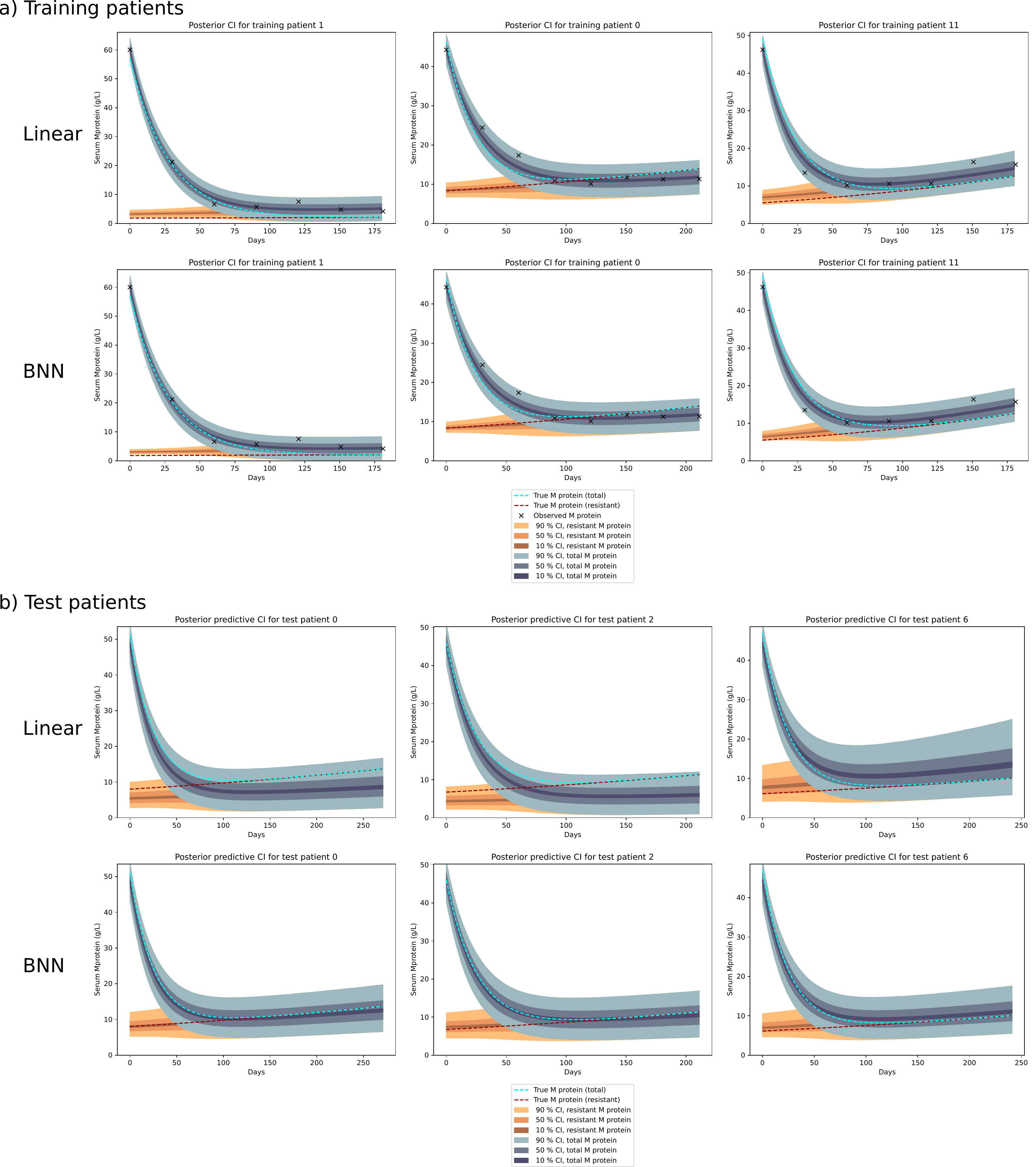}
   \caption{\textbf{Model fit to training data and predictions on test data, with noise standard deviation = 2 and \textit{with} random effects in model parameters of training patients.} a) Model fit of the linear and BNN covariate effect models to patients in the training set, where M protein measurements and baseline covariates were provided to the models.
   b) Predictions on test patients, where only the baseline covariates were provided to the models.}
   \label{fig:train_and_test_5}
\end{figure}
\fi

\end{document}